\documentclass[a4paper]{article}
\usepackage{graphicx}
\usepackage{bm}
\usepackage{amsmath,amssymb}

\title{Application of Video-to-Video Translation Networks to Computational Fluid Dynamics}
\author{Hiromitsu Kigure}
\date{}

\begin{document}
\maketitle

\begin{abstract}
In recent years, the evolution of artificial intelligence, especially deep learning, has been remarkable, and its application to various fields has been growing rapidly. In this paper, I report the results of the application of generative adversarial networks (GANs), specifically video-to-video translation networks, to computational fluid dynamics (CFD) simulations. The purpose of this research is to reduce the computational cost of CFD simulations with GANs. The architecture of GANs in this research is a combination of the image-to-image translation networks (the so-called "pix2pix") and Long Short-Term Memory (LSTM). It is shown that the results of high-cost and high-accuracy simulations (with high-resolution computational grids) can be estimated from those of low-cost and low-accuracy simulations (with low-resolution grids). In particular, the time evolution of density distributions in the cases of a high-resolution grid is reproduced from that in the cases of a low-resolution grid through GANs, and the density inhomogeneity estimated from the image generated by GANs recovers the ground truth with good accuracy. Qualitative and quantitative comparisons of the results of the proposed method with those of several super-resolution algorithms are also presented.
\end{abstract}

\section{Introduction}

Artificial intelligence is advancing rapidly and has come to be comparable to or outperform humans in several tasks. In generic object recognition, deep convolutional neural networks have surpassed human-level performance (e.g,~\cite{He_2015_ICCV, He_2016_CVPR},~\cite{pmlr-v37-ioffe15}). The agent trained by reinforcement learning is capable of reaching a level comparable to professional human game testers (~\cite{Mnih_2015_Nature}). In the case of machine translation, Google's neural machine translation system, using Long Short-Term Memory (LSTM) recurrent neural networks (~\cite{Hochreiter_1997_NC, Gers_2000_NC}), is a typical and famous example and its translation quality is becoming comparable to that of humans (~\cite{2016arXiv160908144W}). 

One of the hottest research topics in artificial intelligence is generative models and one approach to implementing a generative model is generative adversarial networks (GANs) proposed by ~\cite{NIPS2014_5423}. GANs consist of two models trained with conflicting objectives. ~\cite{Radford_2016_ICLR} applied deep convolutional neural networks to those two models, whose architecture is called deep convolutional GANs (DCGAN). DCGAN can generate realistic synthesis images from vectors in the latent space. ~\cite{Isola_2017_CVPR} proposed  the network learning the mapping from an input image to an output image to enable the translation between two images. This network, the so-called pix2pix, can convert black-and-white images into color images, line drawings into photo-realistic images, and so on.

The combination of deep learning and simulation has been recently researched. One of such applications is to use simulation results for improving the prediction performance of deep learning. Since deep learning requires a lot of data for training, numerical simulations that can generate various data by changing physical parameters could help compensate for the lack of training data. Another application is to speed up the solver of computational fluid dynamics (CFD). ~\cite{Guo_2016_KDD} used a convolutional neural network (CNN) to predict velocity fields approximately but fast from the geometric representation of the object. Another example is that velocity fields are predicted from parameters such as source position, inflow speed, and time by CNN (~\cite{DBLP:journals/cgf/KimATKGS19}). Their method is feasible to generate velocity fields up to 700 times faster than simulations. As a more general method, not limited to CFD problems, \cite{raissi2019physics} proposed the physics-informed neural network (PINN), which utilizes a relatively simple deep neural network to find solutions to various types of nonlinear partial differential equations.

GANs also have been combined with numerical simulations to enable a new type of solution method. 
~\cite{DBLP:journals/corr/abs-1709-02432} used the conditional GAN (cGAN) to generate the solution of steady-state heat conduction and incompressible flow from boundary conditions and calculation domain shape/size.  ~\cite{Xie_2018_ACM} proposed a method for super-resolution fluid flow by a temporally coherent generative model (tempoGAN). They showed that tempoGAN can infer high-resolution, temporal, and volumetric physical quantities from those of low-resolution data.

The above-mentioned studies about the combination of GANs and simulations show that GANs can generate the three-dimensional data of the solution of physical equations. The main topic in this research is the translation of images (distributions of the physical quantity) by GANs. In the case that the accuracy of the simulation is particularly important, a large number of computational grids are needed. Additionally, the number of simulation cases for design optimization is typically numerous. It means that the computational cost (machine power and time) becomes large. In such a case, it is important to reduce the computational cost, and one way to do so is to make effective use of low-cost simulations. Based on such an idea, I investigated the feasibility of time-series image-to-image translation: translation from time-series distribution plots in the case of low-resolution computational grids to those in the case of high-resolution grids. A quantitative evaluation of the quality of generated images was also performed.

The method proposed in this paper is the video(sequential images)-to-video translation in which the difference of solutions between the high- and low-resolution grid simulations is learned. Meanwhile, the PINN constructs universal function approximators of physical laws by minimizing the loss function composed of a mismatch of state variables including the initial and boundary conditions and the residual for the partial differential equations (~\cite{MENG2020113250}). In other words, the PINN is an alternative to CFD, while the proposed method is a complement to CFD.

The paper is organized as follows. In section 2, I describe the outline of the simulations whose results are input to GANs and the details of the network architecture. 
In section 3, I give the results of time-series image-to-image translation (in other words, video-to-video translation) of physical quantity distribution and a discussion mainly about the quality of generated images. Conclusions are presented in section 4.

\section{Methods}

\subsection{Numerical Simulations}
I solved the following ideal magnetohydrodynamic (MHD) equations numerically in 2 dimensions to prepare input images to GANs: 
\begin{eqnarray}
\frac{\partial \rho}{\partial t} + \nabla \cdot \left( \rho \bm{v} \right) = 0\\
\frac{\partial}{\partial t}  \left( \rho \bm{v} \right) + \nabla \cdot \left( \rho \bm{v} \bm{v} + p_T \bm{I} - \bm{B} \bm{B} \right) = 0\\
\frac{\partial \bm{B}}{\partial t} + \nabla \cdot \left( \bm{v} \bm{B} - \bm{B} \bm{v} \right) = 0\\
\frac{\partial e}{\partial t} + \nabla \cdot \bigl(\left(e + p_T\right) \bm{v} - \bm{B}  \left(\bm{v} \cdot \bm{B}\right)\bigr) = 0\\
p_T = p + \frac{| \bm{B} |^2}{2}\\
e = \frac{p}{\gamma - 1} + \frac{\rho | \bm{v} |^2}{2} + \frac{| \bm{B} |^2}{2}
\end{eqnarray}
where $\rho, p$, and $\bm{v}$ are the density, pressure, and velocity of the gas; $\bm{B}$ is the magnetic field; $\gamma$ represents the heat capacity ratio and is equal to $5/3$ in this paper; $p_T$ and $e$ represent the total pressure and the internal energy density; $\bm{I}$ is the unit matrix.

One of typical test problems for MHD, the so-called Orszag-Tang vortex problem (~\cite{orszag_tang_1979}), was solved by the Roe scheme (~\cite{ROE1981357}) with MUSCL (monotonic upstream-centered scheme for conservation laws; ~\cite{VANLEER1979101}). 
The initial conditions are summarized in Table \ref{initial_conditions}. $B_0$ is a parameter for controlling the magnetic field strength. The compuational domain is $0 \leq x \leq 1$ and $0 \leq y \leq 1$. The periodic boundary condition is applied in both $x$- and $y$-directions. Simulations for each condition were performed twice on computational grids with different resolutions. The number of grid points is $\left( N_x \times N_y \right) = \left( 51 \times 51\right)$ or $\left( 251 \times 251\right)$. 
In the case of  $\left( N_x \times N_y \right) = \left( 251 \times 251\right)$, the calculation time is more than 70 times longer than in the other case though the obtained solution is expected to be close to the true solution.

\begin{table}[!h]
\caption{The initial conditions of simulations}
\label{initial_conditions}
\centering
\begin{tabular}{|c|c|c|}
\hline
Physical Quantity & Description & value\\
\hline
$\rho$ & density &  $25 \pi/36$ \\
\hline
$v_x$  & x-component of velocity & $-\sin \left( 2 \pi y \right)$\\
\hline
$v_y$  & y-component of velocity & $\sin \left( 2 \pi x \right)$\\
\hline
$v_z$  & z-component of velocity & $0$\\
\hline
$B_x$  & x-component of magnetic field & $-B_0 \sin \left( 2 \pi y \right)$\\
\hline
$B_y$  & y-component of magnetic field & $B_0 \sin \left( 4 \pi x \right)$\\
\hline
$B_z$  & z-component of magnetic field & $0$\\
\hline
$p$ & pressure & $5 \pi/12$\\
\hline
\end{tabular}
\end{table}

\subsection{Generative Adversarial Network Architecture}
After the original concept of GANs was proposed by ~\cite{NIPS2014_5423}, various GANs have been researched. Among such networks, I focused on pix2pix, which is a type of conditional GAN and a network for learning the relationship between the input and output images. The feasibility of translating from the results of low-resolution grid simulations to those of high-resolution grid simulations has been investigated in this research. Furthermore, in order to enable the translation across two time-series, the architecture combined pix2pix and LSTM has been constructed. 

\begin{figure}[!h]
\centering
\includegraphics[width=12.5cm, clip]{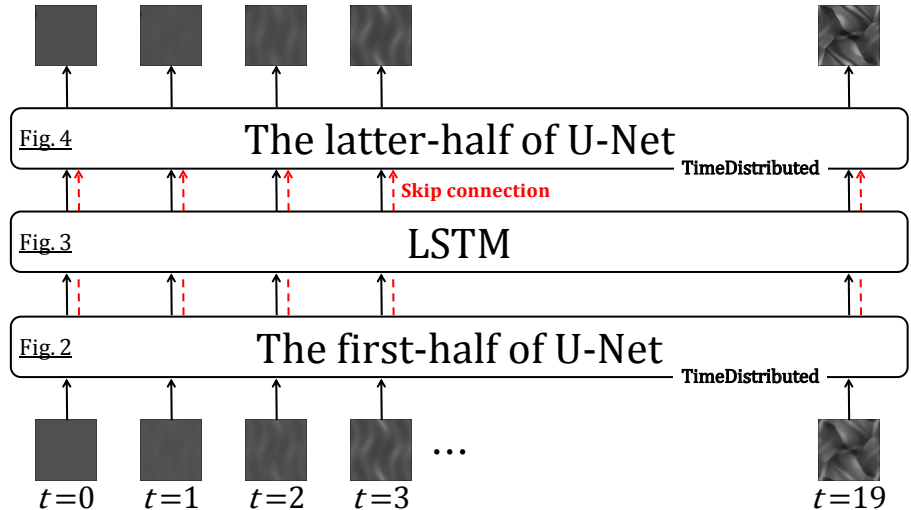}
\caption{Schematic picture of the architecture of the generator in this research. The generator in the original pix2pix network is a U-shaped network (U-Net). In this research, the LSTM layer is inserted into the middle of U-Net. The skip connections from the first-half of U-Net to the latter-half over the LSTM layer are implemented.}
\label{Generator}
\end{figure}

\begin{figure}[!h]
\centering
\includegraphics[width=12.5cm, clip]{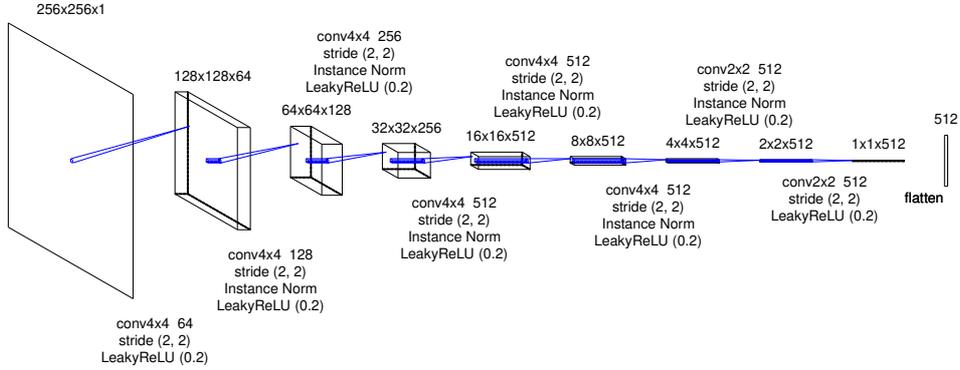}
\caption{The details of the first-half of U-Net. The expression "conv4x4 64" refers to a convolutional layer with a kernel size of $\left( 4 \times 4 \right)$ and $64$ channels. Each feature map is copied and is concatenated to the feature map of the corresponding block in the latter-half of U-Net denoted in Figure \ref{UNet_latter}.}
\label{UNet_first}
\end{figure}

Figure \ref{Generator} shows the schematic picture of the architecture of the generator in this research. 
The role of the LSTM layer is to adjust the image translation dependent on the physical time of the simulation; for the initial state of the simulation $\left( T=0 \right)$, no translation is needed at all, but as physical time passes, progressively larger translations are needed. Note that the weights of the encoder (decoder) before (after) the LSTM layer are the same in the time direction. 
Plots of the time evolution of the density in the low-resolution simulations are input to the generator (plots are read as single-channel images). The input images are converted to vectors by the first-half of a U-shaped network (U-Net). In Figure \ref{UNet_first}, I denote the architecture of the first-half of U-Net in detail. It consists of eight convolutional blocks with a kernel size of $\left( 4 \times 4 \right)$ or $\left( 2 \times 2 \right)$. The instance normalization (~\cite{Ulyanov_2017_CVPR}) is applied except for the first and last blocks. 
The activation function is a leaky rectified linear unit (leaky ReLU; ~\cite{Andrew_LeakyReLU_2013}) with a slope of $0.2$ for all blocks. A 512-dimensional vector is generated at the end of this architecture.

\begin{figure}[!h]
\centering
\includegraphics[width=12.5cm, clip]{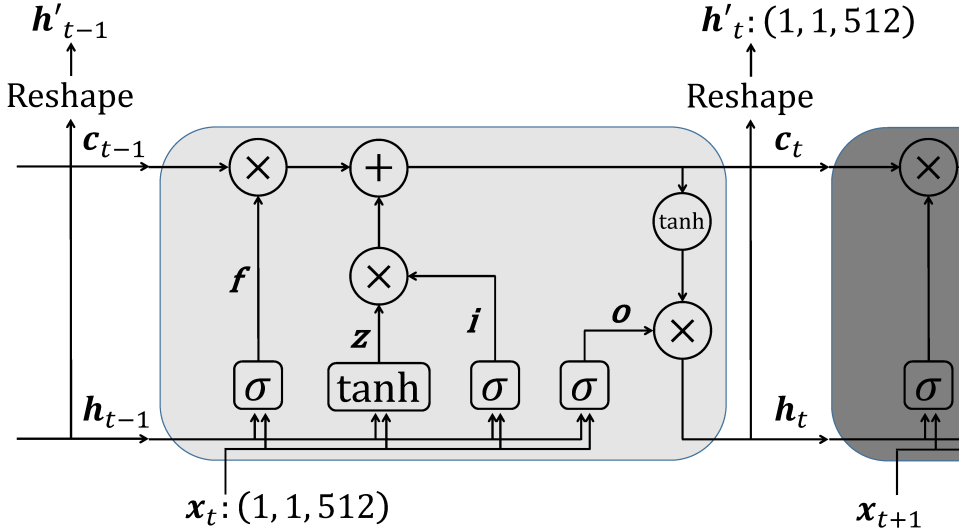}
\caption{The architecture of LSTM. The input to LSTM $\left( \bm{x}_t \right)$ is the vector transformed from an image of density distribution, and the output is the reshaped hidden state vector $\left( \bm{h}^{\prime}_t \right)$ resulting from several operations. The vector $\bm{c}$ is the memory cell, and $\bm{f}$, $\bm{i}$, $\bm{o}$, and $\bm{z}$ are a forget gate, an input gate, an output gate, and part of the term to be added to the memory cell (see equations $\left( 7 \right)$ to $\left( 10 \right)$ for details).}
\label{LSTM}
\end{figure}

A series of 512-dimensional vectors converted from the time-series plots is input to the LSTM layer. An input vector $\bm{x}_t$ originated from the plot at time $=t$ is calculated with the hidden state $ \bm{h}_{t-1}$ and memory cell $\bm{c}_{t-1}$. A forget gate $\left( \bm{f} \right)$, an input gate $\left( \bm{i} \right)$, an output gate $\left( \bm{o} \right)$, and part of the term to be added to the memory cell $\left( \bm{z} \right)$ in Figure \ref{LSTM} are calculated as follows:
\begin{eqnarray}
\bm{f} &=& \sigma \left( \bm{W}_f \bm{x}_t + \bm{R}_f \bm{h}_{t-1} + \bm{b}_f \right)\\
\bm{i} &=& \sigma \left( \bm{W}_i \bm{x}_t + \bm{R}_i \bm{h}_{t-1} + \bm{b}_i \right)\\
\bm{o} &=& \sigma \left( \bm{W}_o \bm{x}_t + \bm{R}_o \bm{h}_{t-1} + \bm{b}_o \right)\\
\bm{z} &=& \tanh \left( \bm{W}_z \bm{x}_t + \bm{R}_z \bm{h}_{t-1} + \bm{b}_z \right)
\end{eqnarray}
where $\sigma$ is the sigmoid function and $\tanh$ is the hyperbolic tangent function; $\bm{W}_{\cdot}$ and $\bm{R}_{\cdot}$ are the input-to-hidden weight matrices and the recurrent weight matrices; $\bm{b}_{\cdot}$ are bias vectors.
The hidden state and memory cell are updated by: 
\begin{eqnarray}
\bm{c}_t &=& \bm{f} \odot \bm{c}_{t-1} + \bm{i} \odot \bm{z}\\
\bm{h}_t &=& \bm{o} \odot \tanh \left( \bm{c}_t \right)
\end{eqnarray}
The hidden state $\bm{h}_t$ is reshaped as $\left( 1, 1, 512 \right)$.

\begin{figure}[!h]
\centering
\includegraphics[width=12.5cm, clip]{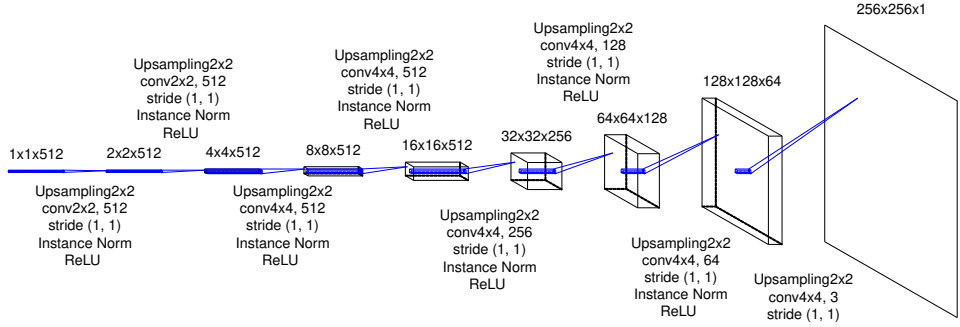}
\caption{The details of the latter-half of U-Net. The expression "Upsampling2x2" refers to an upsampling layer that doubles the size of input by copying one value twice horizontally and vertically, respectively. From the first-half of U-Net displayed in Figure \ref{UNet_first}, feature maps are passed to corresponding blocks and are concatenated to the feature maps output from the previous blocks.}
\label{UNet_latter}
\end{figure}

The reshaped hidden state $\bm{h}_t^{\prime}$ is passed to the latter-half of U-Net and is decoded to the image data (see Figure \ref{UNet_latter}). This part consists of eight deconvolutional blocks with an upsampling of the feature map, convolution with a kernel size of $\left( 2 \times 2 \right)$ or $\left( 4 \times 4 \right)$ (the size of the feature map does not change because the stride of convolution is 1), the instance normalization and activation by ReLU function except the last block. As seen in Figure \ref{Generator}, the generator outputs synthetic time-series plots of density distribution.

\begin{figure}[!h]
\centering
\includegraphics[width=12.5cm, clip]{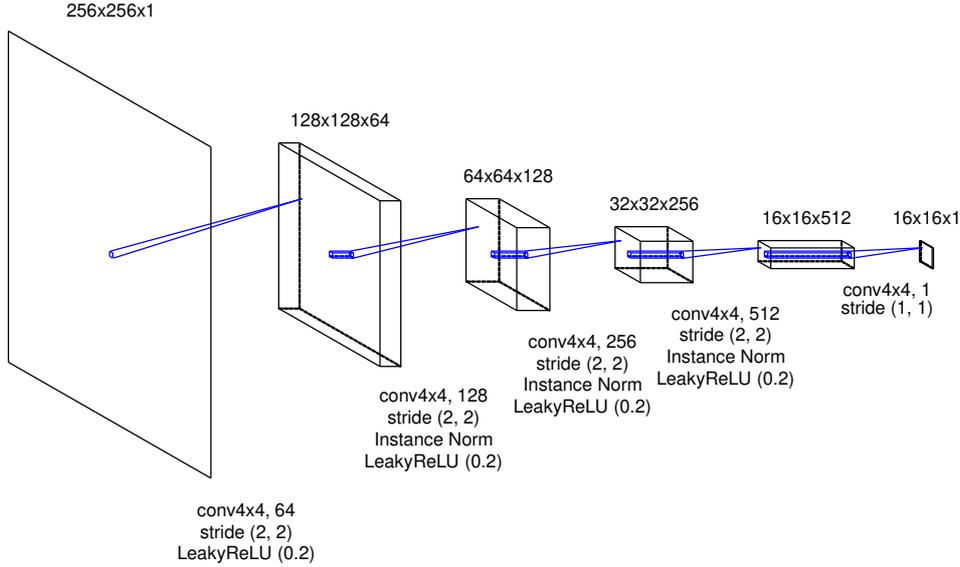}
\caption{The details of the architecture of the discriminator in this research.}
\label{Discriminator}
\end{figure}

The authenticity of the images is judged by the discriminator. Figure \ref{Discriminator} shows the details of the architecture of the discriminator in this research. A real image (plot of the density distribution in a high-resolution simulation) or a synthesis image is input to the discriminator. It consists of five convolutional blocks with a kernel size of $\left( 4 \times 4 \right)$. The instance normalization is applied except for the first and last blocks. Except for the last block, the leaky ReLU function with a slope of $0.2$ is applied as the activation function. The $16 \times 16$ patch is eventually output. The discriminator classifies each patch into real or synthetic. We call its architecture the patchGAN (~\cite{Isola_2017_CVPR}).

The objective of the network is the same as the regular pix2pix as follows:
\begin{eqnarray}
G^{\ast} &=& \arg \, \underset{G}{\min} \, \underset{D}{\max} \, \mathcal{L}_{cGAN} \left( G, D \right) + \lambda \mathcal{L}_{L1} \left( G \right)\\
\mathcal{L}_{cGAN} \left( G, D \right) &=& \mathbb{E} \left[ \log D \right( x, y\left) \right] + \mathbb{E} \left[ \log \left( 1 - D \left( x, G \left( x \right) \right) \right) \right]\\
\mathcal{L}_{L1} \left( G \right) &=& \mathbb{E} \left[ \| y - G \right( x \left) \| \right]
\end{eqnarray}
where $G$ and $D$ denote the generator and discriminator, $\lambda$ is the weighted sum parameter and equal to 100 in this research, and $x$ and $y$ mean the source and target images. $G \left( x \right)$ returns a synthesis image and $D \left( x, y \right)$ or $D \left( x, G \left( x \right) \right)$ returns the probability that $y$ or $G \left( x \right)$ is a real target image. $\mathcal{L}_{L1} \left( G \right)$ is the mean absolute error (L1 loss) calculated from the pixel-wise comparison between the real image and the synthetic image.
The optimizer is Adam with a learning rate of 0.0002.

The architecture is implemented using Keras 2.5.0 and TensorFlow as a backend. The model was trained on Google Colaboratory with Tesla P100-PCIe GPU. For applying the convolution and deconvolution to the sequential data, sets of operations as shown in Figures \ref{UNet_first}, \ref{UNet_latter}, and \ref{Discriminator} are passed to the TimeDistributed layer. The skip connections are implemented by feeding the outputs of the previous upsampling block in the latter-half of U-Net and the same-level (it means that the size of the feature map is the same) convolutional block in the first-half of U-Net to the Concatenate layer. The 'return\_sequences' and 'stateful' parameters in the LSTM layer are set to True and False, respectively.

\section{Results and Discussion}
In this chapter, I show the results of time-series image-to-image translation for the training datasets first and then explain the way to evaluate the quality of the synthesis images quantitatively. The evaluation result of the synthesis images for the training datasets is presented next. 
Then, I show the results for the testing datasets. Finally, the quality of the synthesis images is compared with those of images upsampled by conventional super-resolution algorithms. The conditions (the magnetic field strength) of the simulations are shown in Table \ref{parameter_simulations} that summarizes the details of the training and testing datasets. The sixteen cases were performed to prepare the training datasets, and the nineteen cases were performed to prepare the testing datasets. For each case, two simulations were run with the high-resolution and the low-resolution grids.

\begin{table}[!h]
\caption{The details of the training and testing datasets}
\label{parameter_simulations}
\centering
\scalebox{0.7}{
\begin{tabular}{|c|c|c|c|c|}
\hline
\begin{tabular}{c}
Training/\\Testing
\end{tabular}
&
\begin{tabular}{c}
The number\\of cases
\end{tabular}
&
\begin{tabular}{c}
The total number\\of images
\end{tabular}
&
\begin{tabular}{c}
The pixel size\\of images
\end{tabular}
&The value of $\bm{B}_0$ \\
\hline
Training & 16 & 320 & $256 \times 256$ &
\begin{tabular}{c}
0.1 -- 1.5 with the interval\\of 0.1, and 2.0
\end{tabular}
\\
\hline
Testing & 19 & 380 & $256 \times 256$ &
\begin{tabular}{c}
0.15 -- 1.55 and 1.6 -- 1.9 \\with the interval of 0.1
\end{tabular}
\\
\hline
\end{tabular}
}
\end{table}

\subsection{Results for the Training Datasets}
\begin{figure}[!h]
\centering
\includegraphics[width=12cm, clip]{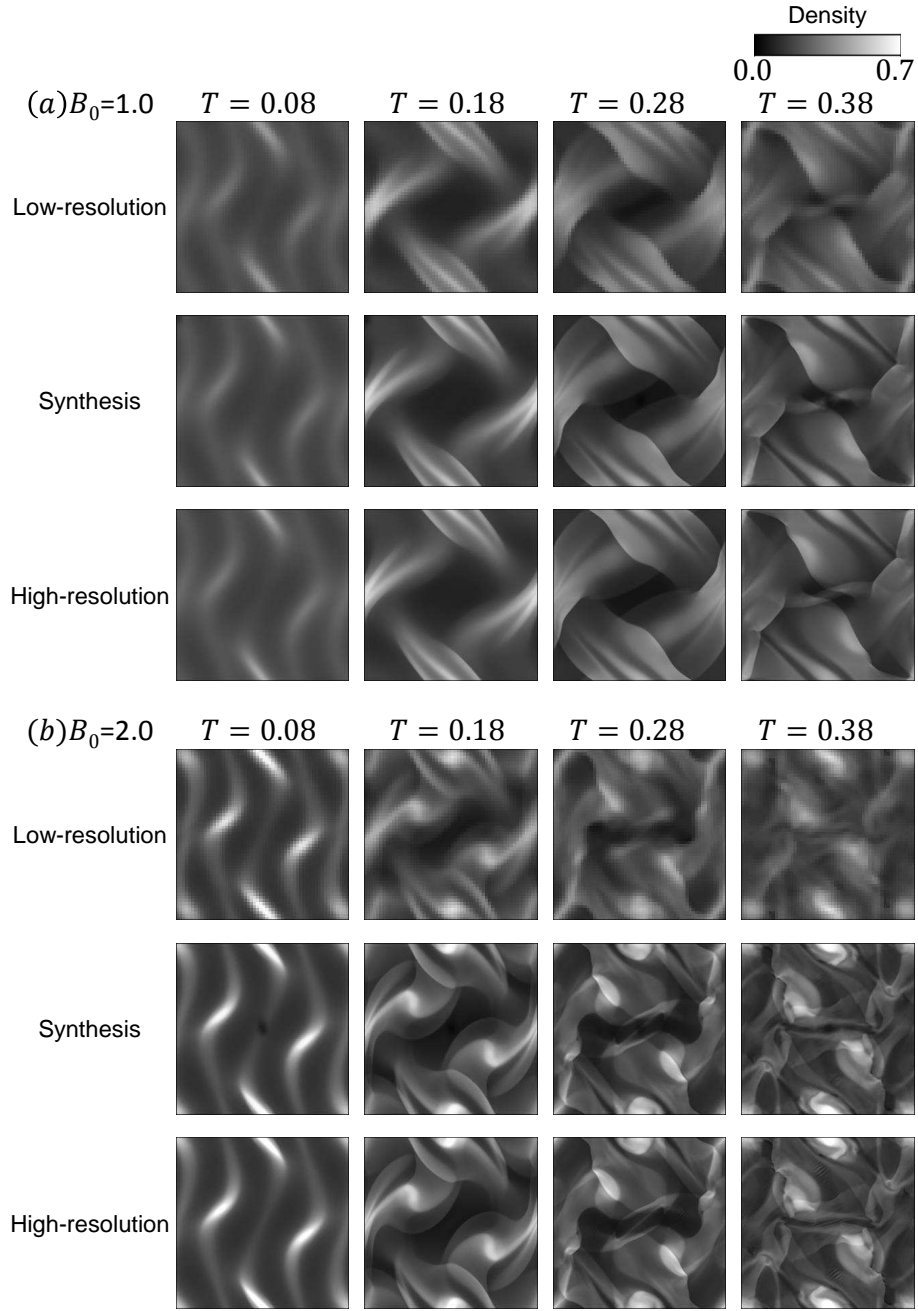}
\caption{Two examples of the time-evolution of density distribution for the training datasets.}
\label{Density_Distribution_train}
\end{figure}

\begin{figure}[!h]
\centering
\includegraphics[width=12cm, clip]{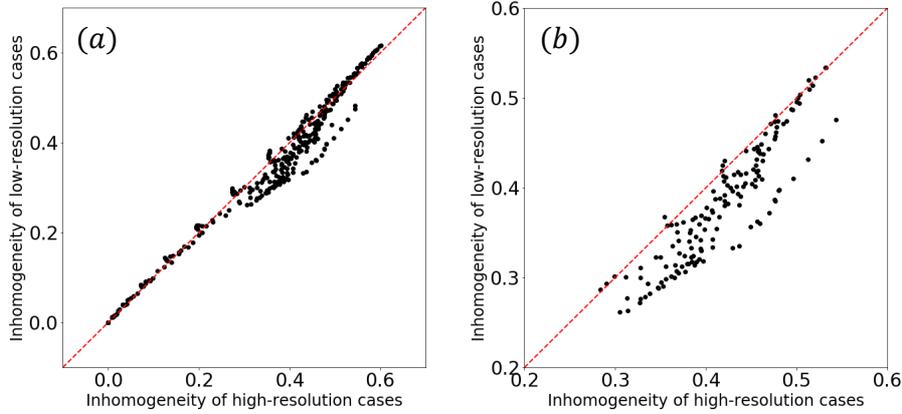}
\caption{Comparison of the inhomogeneity of the density between the high-resolution grid cases and the low-resolution grid cases for the training datasets. (a)The inhomogeneities for all time-series and all magnetic field strength cases are plotted. (b)The inhomogeneities for $T \geq 0.12$ and $B_0 \geq 0.6$ are plotted.}
\label{inho_high_low}
\end{figure}

Figure \ref{Density_Distribution_train} shows two examples of the time-evolution of density distribution for the training datasets. The top and bottom images of Figure \ref{Density_Distribution_train}-(a), (b) show the simulation results, and 
the middle images are synthesis ones generated from the top ones (the results of low-resolution grid simulations) through the generator. Compared to the high-resolution grid cases, the density distributions in the low-resolution grid  cases show less fine structure and become closer to the uniform. Figure \ref{inho_high_low} displays the comparison of the inhomogeneity of the density between the high-resolution grid cases and the low-resolution grid cases. The inhomogeneity is defined by $\alpha = \sigma_\rho / \bar{\rho}$, where $\sigma_\rho$ and $\bar{\rho}$ are the standard deviation and the average of the density. In the low-resolution grid, the numerical diffusion is larger than in the high-resolution grid, and therefore the inhomogeneity of the density tends to be smaller especially from the middle stage of the vortex development and in the relatively strong magnetic field (see Figure \ref{inho_high_low}-(b)). The synthesis images reproduce the fine structures of the density distributions and appear to be well consistent with the high-resolution grid results.

\begin{figure}[!h]
\centering
\includegraphics[width=7cm, clip]{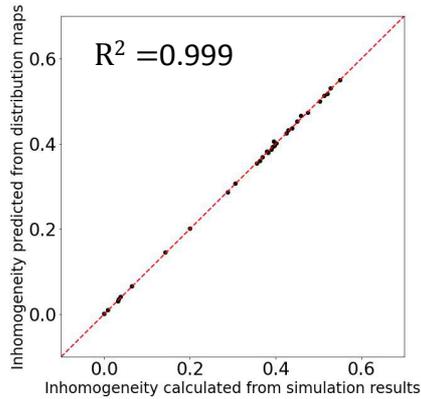}
\caption{Comparison of the inhomogeneity calculated from the density values on the grids and the inhomogeneity predicted from the distribution maps. The coefficient of determination $\left( R^2 \right)$ is equal to 0.999.}
\label{MLP_evaluation}
\end{figure}

To quantitatively evaluate the quality of the synthesis images, I estimated the density inhomogeneity from the distribution map. When calculating the density inhomogeneity from the simulation result, we can use the value of the density on each grid; however, the density distribution maps (including synthesis images in this research) have only the information of the RGB values. Therefore, to estimate the density inhomogeneity from the distribution map, I trained a three-layer fully connected neural network with 196608 (256pixel $\times$ 256pixel $\times$ 3) inputs, two hidden layers of 1024 and 128 neurons and one output layer. Figure \ref{MLP_evaluation} shows the result of the inhomogeneity prediction from the density distribution maps. The horizontal axis is the inhomogeneity calculated from the density values on the grids, and the vertical axis is the inhomogeneity predicted from the distribution maps by the trained neural network. The coefficient of determination $\left( R^2 \right)$ is equal to 0.999. Thus we conclude that the trained neural network provides an accurate estimation of the density inhomogeneity from the distribution maps and the synthesis images.

\begin{figure}[!h]
\centering
\includegraphics[width=7cm, clip]{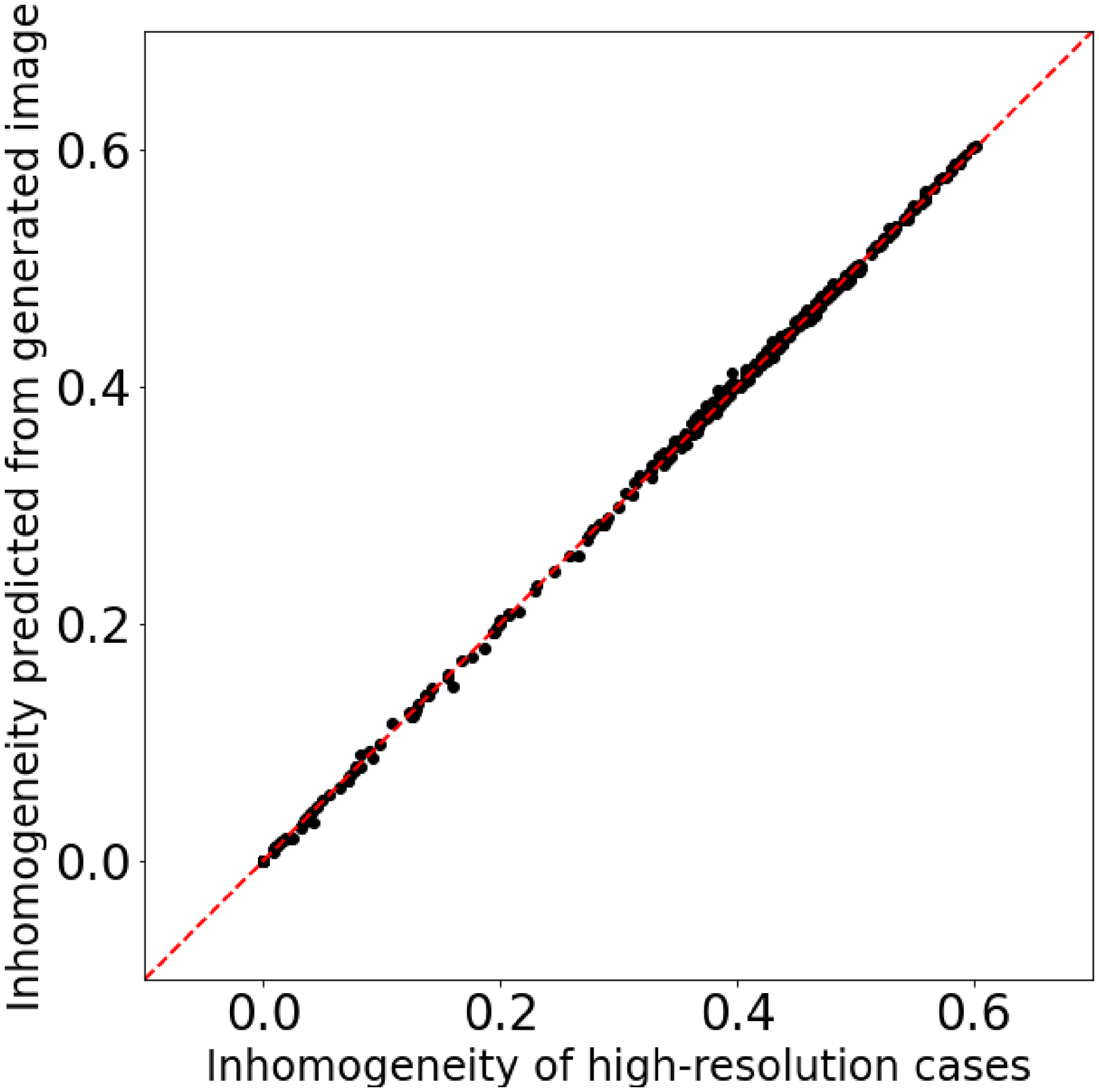}
\caption{Comparison of the inhomogeneity of the high-resolution grid simulation results and the inhomogeneity predicted from the synthesis images for the training datasets.}
\label{training_evaluation}
\end{figure}

We can quantitatively evaluate the quality of the synthesis images by inputting those into the neural network and comparing the outputted inhomogeneity with the inhomogeneity calculated from the high-resolution grid simulation results. 
Figure \ref{training_evaluation} shows that the inhomogeneity predicted from the synthesis images matches that calculated from the high-resolution grid simulation results with good accuracy; therefore the quality of the synthesis images is definitely good for the training datasets.

\begin{figure}[!h]
\centering
\includegraphics[width=12cm, clip]{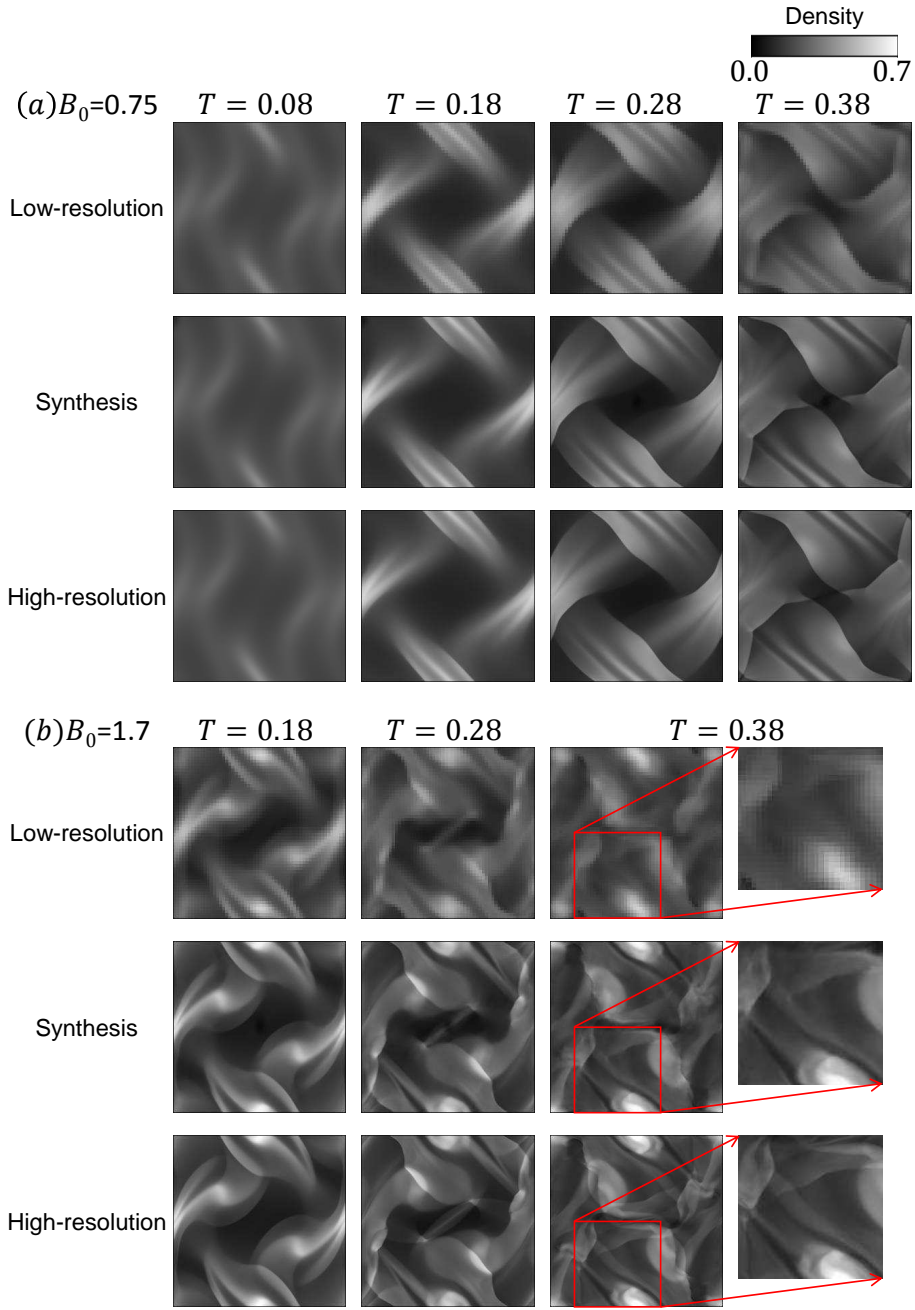}
\caption{Examples of the time-evolution of density distribution for the testing datasets.}
\label{Density_Distribution_test}
\end{figure}

\begin{figure}[!h]
\centering
\includegraphics[width=6cm, clip]{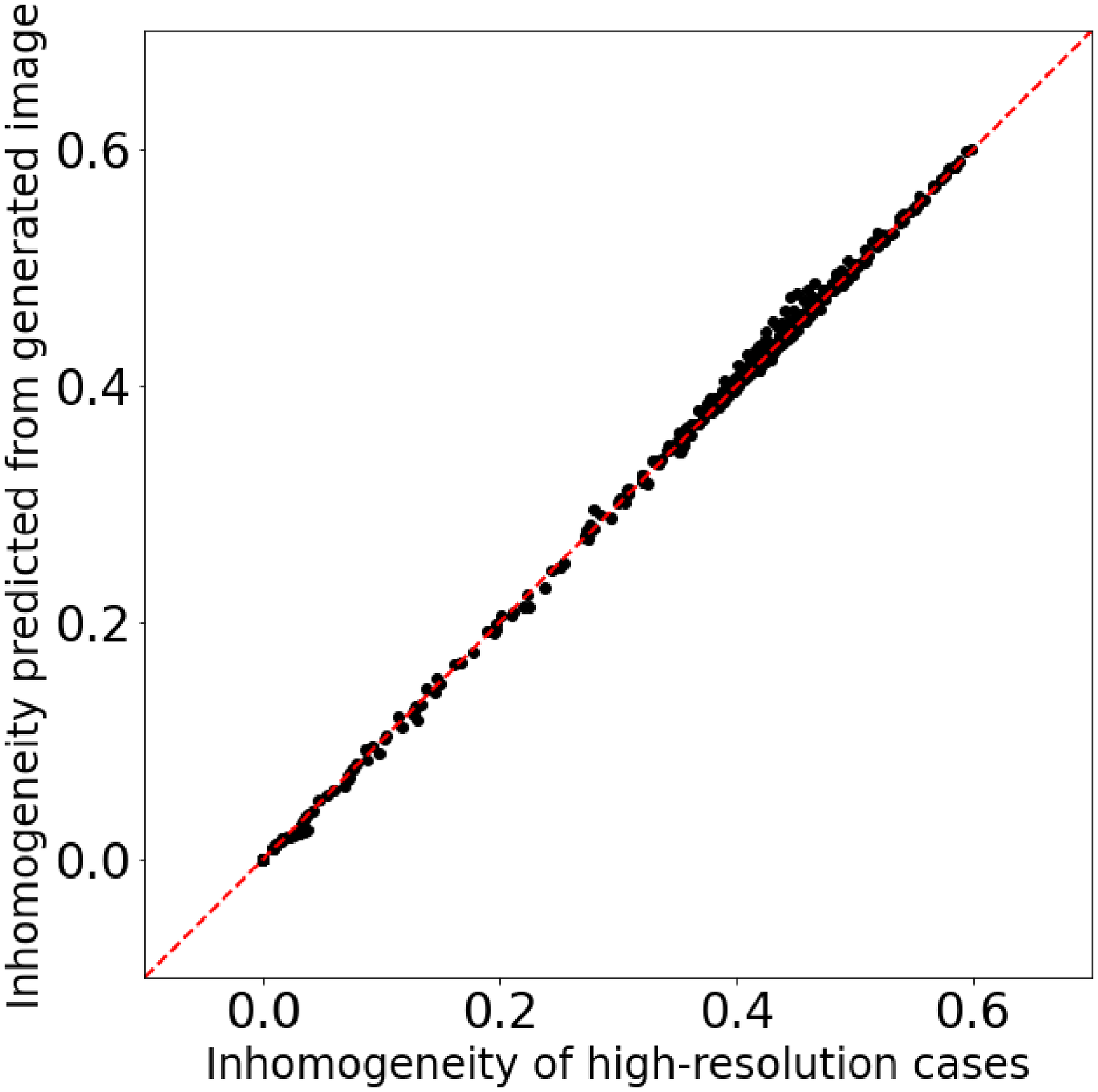}
\caption{Comparison of the inhomogeneity of the high-resolution grid simulation results and the inhomogeneity predicted from the synthesis images for the testing datasets.}
\label{testing_evaluation}
\end{figure}

\subsection{Results for the Testing Datasets}
In the previous subsection, I have shown that the results for the training datasets are pretty good. However, the generalization ability needs to be investigated for practical use. The testing datasets (the magnetic field strength is different from the training datasets as shown in Table \ref{parameter_simulations}) that were not used for training are input to the trained model, and the synthesis images are output from the generator. Figures \ref{Density_Distribution_test}-(a), (b) show the comparison of the simulation results and the synthesis images for two example cases. From the 19 cases in the testing datasets, the results for the cases with $B_0=0.75$ and $1.7$ were selected for presentation. The $B_0=1.7$ case is especially suitable for verifying the generalization ability because there is no training data between $B_0=1.5$ and $2.0$. The top images show the time evolution of the density distribution of low-resolution grid simulations, which are input for the generator; the bottom images show that of high-resolution grid simulations, which are compared with the synthesis images; the middle images are synthesis ones generated through the generator. As with the cases for the training datasets, the synthesis images qualitatively reproduce the density distributions of the high-resolution grid simulations. 
Even in the $B_0=1.7$ case, the synthesis images show the fine structure of the density distribution very similar to that in the ground truth images, as shown in the zoomed-in image in Figure \ref{Density_Distribution_test}-(b).

Figure \ref{testing_evaluation} is almost the same as Figure \ref{training_evaluation} but for the testing datasets. The density inhomogeneity predicted from the synthesis images through the fully connected neural network (explained in the previous subsection) is in good agreement with the inhomogeneity calculated from the results of high-resolution grid simulations. This result indicates that the method in this research is capable of obtaining high generalization ability.

\subsection{Comparison with conventional super-resolution algorithms}
To demonstrate the effectiveness of the proposed method and the quality of the generated images, I compare the results with those obtained by conventional super-resolution algorithms. The algorithms investigated here are a bicubic interpolation, a Lanczos interpolation, and Laplacian Pyramid Super-Resolution Network (LapSRN; ~\cite{Lai_2017_CVPR}). The pixel size of the image to be used as the basis of the super-resolution is $64 \times 64$, and each algorithm quadruples the pixel size. These results were compared qualitatively and quantitatively 
with the result of high-resolution grid simulation and the image generated by the proposed method. Plots of the density distribution in high-resolution simulations in the training datasets were used to train LapSRN. 

\begin{figure}[!h]
\centering
\includegraphics[width=14cm, clip]{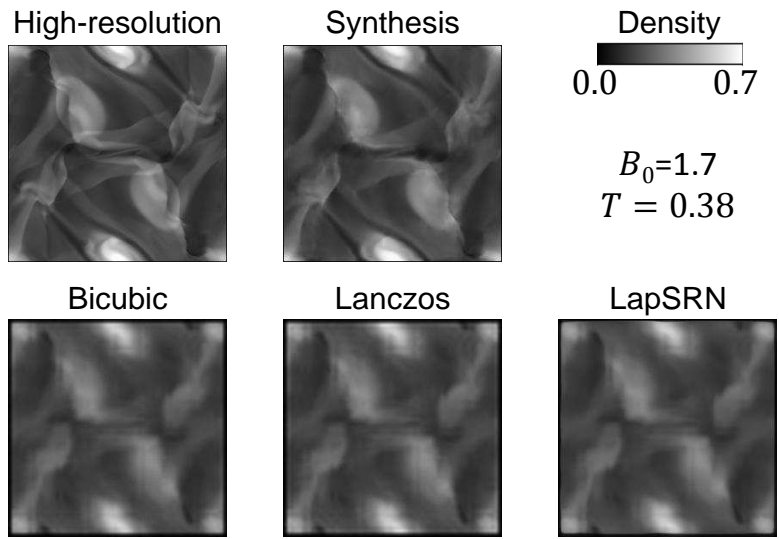}
\caption{Comparison of the results of the conventional super-resolution algorithms with that of the proposed method and ground truth.}
\label{comparison_SR_image}
\end{figure}

\begin{figure}[!h]
\centering
\includegraphics[width=12cm, clip]{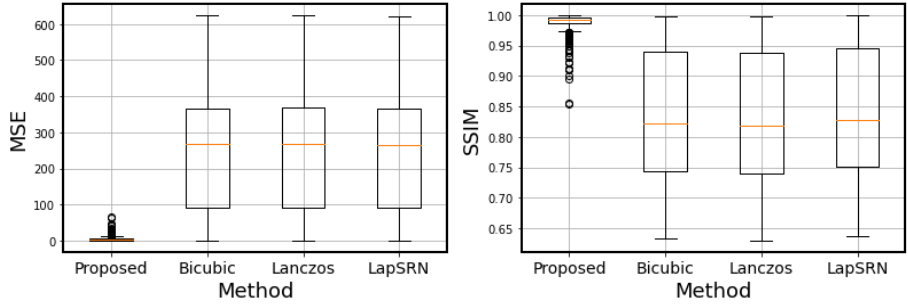}
\caption{Box plots of the pixel-wise mean squared error (MSE) and the structural similarity index measure (SSIM) calculated in the testing datasets (380 images).}
\label{MSE_SSIM}
\end{figure}

I performed the super-resolution algorithms to the testing datasets (380 images). As an example, the results for the $B_0 = 1.7$ and $T = 0.38$ case are compared in Figure \ref{comparison_SR_image}. In this case, none of the three conventional super-resolution algorithms can work with a quality comparable to the method proposed in this research. To compare the proposed method with the others quantitatively, the pixel-wise mean squared error (MSE) and the structural similarity index measure (SSIM; \cite{1284395}) are calculated between the ground truth image and the synthesis image or the result of super-resolution. Figure \ref{MSE_SSIM} shows that the quality of the synthesis images by the proposed method is significantly high compared to that of the results by the conventional super-resolution algorithms.

\subsection{Application of this research}
In this subsection, I discuss an application of this research. As mentioned above, results of high computational cost simulations can be estimated from those of low-cost simulations by the method in this paper. However, it is important to note that simulation results of quite a few cases are needed to train the network\footnote{In this research, simulation results of 16 cases were used as the training datasets; the training was successful with a relatively small number of data, probably due to the simple situation. If the target is a simulation of a realistic engineering situation, it is expected that much more data will be needed for the training.}. 
Therefore, it is not beneficial for a small number of simulations. The more simulations are required, the greater the benefits arise. One such case is optimization based on CFD simulations. As the number of objective variables to be optimized increases, the number of calculations required to obtain the desired performance is expected to increase; in some cases, it takes several thousand cases to evaluate. In such multi-objective optimization simulations, for example, the first dozens to several hundred cases are simulated on both high- and low-resolution grids, and the results are used to train the GANs. After the GANs are trained, low-resolution grid simulations are run, the results are input to the GANs to reproduce the results of high-resolution grid simulations, and objective variables are estimated from synthesis images by, for example, a neural network.

I demonstrate the estimation of computational cost reduction. If the number of simulations required originally and that to train the GANs are $N$ (several thousands in some cases) and $N_t$ ($N > N_t$), the calculation times of the high- and low-resolution grid simulations are $T_h$ and $T_l$ ($T_h > T_l$), and the computational cost to train the GANs is $T_t$, the computational cost reduction is roughly equal to 
\begin{eqnarray}
N \times T_h - \left( N_t \times T_h + T_t + N \times T_l \right)
\end{eqnarray}
where the first term corresponds to the computational cost in the case that all simulations are run on the high-resolution grid, and the second term corresponds to that in the case that the method in this research is applied (the cost to reproduce the results of high-resolution grid simulations by the GANs is negligible compare to performing the simulations). In this way, by substituting low-resolution grid simulations and the result conversion by the GANs for quite a part of high-resolution grid simulations, a great reduction of the computational cost should be achieved.

\section{Conclusions}
In this paper, I validated an idea to use GANs for reducing the computational cost of CFD simulations. I studied the idea of reproducing the results of high-resolution grid simulations with a high computational cost from those of low-resolution grid simulations with a low computational cost. More specifically speaking, distribution maps of a physical quantity in time series were reproduced using pix2pix and LSTM. The quality of the reproduced synthesis images was good for both the training and testing datasets. The conditions treated in this paper are simple; the computational region is a square with a constant grid interval, the boundary conditions are cyclic, and the governing equations are the ideal MHD equations. In the next step, I need to examine the idea in more realistic conditions.

\bibliographystyle{plain}
\bibliography{kigure_2021}

\end{document}